**The epigenome of evolving Drosophila neo-sex chromosomes: dosage compensation and heterochromatin formation**

Qi Zhou[1]\*, Christopher E. Ellison[1]\*, Vera B. Kaiser[1], Artyom A. Alekseyenko[2], Andrey A. Gorchakov[2,3], Doris Bachtrog[1]#

[1]Department of Integrative Biology, University of California Berkeley, Berkeley, CA 94720

[2]Department of Genetics, Harvard Medical School, Boston, MA 02115

[3]Laboratory of Chromosome Engineering, Institute of Molecular and Cellular Biology, Novosibirsk, Russia 630090

\* these authors contributed equally,

# corresponding author: dbachtrog@berkeley.edu

**Abstract**

**Sex chromosomes originated from autosomes but have evolved a highly specialized chromatin structure. Drosophila Y chromosomes are composed entirely of silent heterochromatin, while male X chromosomes have highly accessible chromatin and are hypertranscribed as a result of dosage compensation. Here, we dissect the molecular mechanisms and functional pressures driving heterochromatin formation and dosage compensation of the recently formed neo-sex chromosomes of *Drosophila miranda*. We show that the onset of heterochromatin formation on the neo-Y is triggered by an accumulation of repetitive DNA. The neo-X has evolved partial dosage compensation and we find that diverse mutational paths have been utilized to establish several dozen novel binding consensus motifs for the dosage compensation complex on the neo-X, including simple point mutations at pre-binding sites, insertion and deletion mutations, microsatellite expansions, or tandem amplification of weak binding sites. Spreading of these silencing or activating chromatin modifications to adjacent regions results in massive mis-expression of neo-sex linked genes, and little correspondence between functionality of genes and their silencing on the neo-Y or dosage compensation on the neo-X. Intriguingly, the genomic regions being targeted by the dosage compensation complex on the neo-X and those becoming heterochromatic on the neo-Y show little overlap, possibly reflecting different propensities along the ancestral chromosome that formed the sex chromosome to adopt active or repressive chromatin configurations. Our findings have broad**

implications for current models of sex chromosome evolution, and demonstrate how mechanistic constraints can limit evolutionary adaptations. Our study also highlights how evolution can follow predictable genetic trajectories, by repeatedly acquiring the same 21-bp consensus motif for recruitment of the dosage compensation complex, yet utilizing a diverse array of random mutational changes to attain the same phenotypic outcome.

## Introduction

Sex chromosomes evolve from ordinary autosomes [1]. Degeneration of the Y chromosome is a general facet of sex chromosome evolution, and old Y chromosomes are gene poor, often contain high amounts of repetitive DNA, and in Drosophila the Y is entirely heterochromatic [2]. The euchromatic, gene-rich X, in contrast, has adopted a hyperactive chromatin configuration, resulting in hyper-transcription of X-linked genes in male Drosophila (i.e. dosage compensation). While ultimately resulting in opposite phenotypic outcomes, the formation of hyperactive chromatin on the X and silent heterochromatin on the Y has intriguing parallels [3,4]. Both are initiated at specific nucleation sites, are associated with characteristic histone modifications, and spreading of the modified chromatin configuration across tens of kilobases allows genomic neighborhoods to adopt a similar silent or hyperactive chromatin state.

In particular, dosage compensation in *Drosophila* occurs by doubling the transcription rate of X-linked genes in males [5], through recruitment of the MSL-complex to specific chromatin entry sites (CES) on the X in a sequence-specific manner [6,7]. The MSL-complex in *D. melanogaster* targets a 21-bp GA-rich DNA segment found at most CES, termed the MSL recognition element (MRE) [6,7], and roughly 150 CES have been identified on the X chromosome of *D. melanogaster* [6,7]. Co-transcriptional targeting and spreading of the MSL-complex along the X chromosome results in MSL-binding of most active genes and their transcriptional upregulation, mediated by changes in the chromatin structure of the X (through H4K16 acetylation H4K16ac [8-10]). Less is known about how a genomic region is targeted to adopt a heterochromatic configuration, but repetitive elements are thought to be involved in triggering the initiation and spreading of silencing heterochromatin [11,12]. Several studies, particularly in yeast, have suggested that RNAi-mediated silencing pathways can initiate the formation of heterochromatin (reviewed in [13-15]). Transcripts from repetitive elements in the centromeric region of fission yeast are processed into small interfering siRNAs and incorporated into a RNAi-induced transcriptional silencing complex that recognizes and binds homologous regions to initiate gene silencing via H3K9 methylation [16,17]. Drosophila centromeres also contain actively transcribed satellite- and transposon-fragment repeats ([18,19]) and mutations in genes encoding the RNAi pathway disrupt HP1 localization and heterochromatin formation. This suggests that a similar mechanism for RNAi-mediated heterochromatin assembly operates in Drosophila as well, and recent work has shown how transposons and the piRNA pathway affect chromatin patterns in Drosophila [20-22].

How epigenetic modifications are acquired on sex chromosomes is a puzzle, and little is known about how dosage compensation and heterochromatin formation evolve on a newly formed sex chromosome pair. That is, how are new nucleation sites to trigger dosage compensation or heterochromatin formation acquired on a former autosome, how does a genomic region become targeted to adopt a hypertranscribed or heterochromatic appearance, what functional pressures drive the evolution of dosage compensation and heterochromatin formation, and how do they interact?

In *D. miranda*, a new sex chromosome formed about 1MY ago, through a fusion of an autosome with the ancestral Y chromosome [23] (**Figure** 1A). These 'neo-sex' chromosomes are at an intermediate stage in the transition from a pair of autosomes to a pair of heteromorphic sex chromosomes. In particular, the neo-Y of *D. miranda* is undergoing massive degeneration: it is rapidly accumulating repetitive DNA, is evolving a heterochromatic appearance and about 40% of its ancestral genes have become non-functional (i.e. they have acquired frame shift mutations or stop codons on the neo-Y, or have been completely lost, [24-27]). Gene expression is generally reduced at neo-Y genes compared to their neo-X homologs [25,28] but its chromatin structure, and the association of heterochromatin and gene expression, has not yet been studied at the molecular level. The neo-X, in contrast, is beginning to acquire partial dosage compensation by coopting the MSL machinery that has evolved to compensate the ancestral X chromosome that is shared among all Drosophila species [29-31] (**Figure** 1B-D), and MSL-mediated dosage compensation is found throughout the Drosophila genus [29-31]. Several components of the dosage compensation complex have been shown to have male-specific gene expression patterns and target the newly formed X chromosomes of *D. miranda* males ([29,30], **Supplementary Figure 1**), and the characteristic histone modification induced by the MSL-complex (H4K16ac) is also enriched at all the male X chromosomes, including the neo-X [30,31], supporting that the function of MSL is conserved across the Drosophila genus. We have previously studied MSL-binding patterns in *D. miranda* using ChIP-seq, in order to identify genes on the X chromosomes that are targeted by the dosage compensation complex, and showed that the sequence motif and function of CES is conserved between *D. miranda* and *D. melanogaster* [32]. ChIP-seq profiling of MSL3 identified 68 novel CES that have already evolved on the neo-X of *D. miranda*, and, via spreading of the MSL-complex, about 607 neo-X genes (22% of all annotated genes on the neo-X) are MSL-bound (and 37% of actively transcribed genes) [32]. Binding of the MSL-complex may be more transient for some genes [33], and about 1203 genes on the neo-X are bound by MSL and/or enriched for H4K16ac, the histone mark deposited by the MSL-complex (i.e. 44% of all genes, and 73% of actively transcribed genes may already be dosage compensated on the neo-X; [32]). The mutational paths that create novel CES on the neo-X, and the dynamic interactions of evolving dosage compensation on the neo-X versus degeneration of the neo-Y, however, have not been systematically investigated. Here, we examine the acquisition of dosage compensation on the neo-X chromosome and formation of heterochromatin on the neo-Y of *D. miranda* at the molecular and functional level, in order to identify how epigenetic modifications evolve on nascent sex chromosomes.

**Results**

**Acquisition of MSL-binding sites on the neo-X**. Previous work has shown that parts of the neo-X of *D. miranda* have acquired MSL-mediated dosage compensation [29-32], however, the evolutionary processes involved in the formation of the CES on this chromosome remain unknown. Evolving novel CES along a new X chromosome to initiate dosage compensation presents a challenge. To recruit the MSL complex, a 21-bp GA-rich DNA segment (the MSL recognition element or MRE, see **Figure** 2A) [6,7], needs to be acquired on many locations along the newly formed X. Multiple mutations may be necessary to evolve that sequence motif at a particular genomic location and the emergence of a novel binding site may require the presence of a pre-site on the neo-X (i.e. a site that shows high sequence similarity to a MRE). While our previous study provided evidence that a CES can be created by tandem amplification of a short GA-rich sequence [32], that work focused on a single CES and it remains unclear whether the other 67 CES evolved via similar mechanisms. The neo-X chromosome of *D. miranda* segregates as an autosome in its closely related sister species *D. pseudoobscura* and *D. affinis*, and comparative sequence analysis allows us to reconstruct, to some extent, the path evolution has taken to acquire the MSL-binding motifs on the neo-X. In particular, we aimed to identify mutational events within the 68 putative CES on the neo-X that were unique to *D. miranda* and would create a novel or stronger MRE on the neo-X (**Figure** 2B). We excluded 2 CES regions that we were not able to align to *D. affinis*. In 25 cases, we were not able to identify the mutations that created a putative CES on the neo-X (see **Figure** 2B). For the remaining 41 CES, we found several different mutational routes to evolve novel MREs on the neo-X (see **Figure** 2C for representative examples). At 28 CES, insertion or deletion mutations created a novel CES on the neo-X at a genomic region that has little or no affinity for the MSL complex in outgroup species. In 7 cases, simple nucleotide mutations have generated a stronger recruitment motif for the MSL-complex on the neo-X at a pre-site, and in 4 cases, a GA-microsatellite expansion created a stronger MRE motif at a pre-site. Another mechanism to generate new CES, found twice, involves the modification and tandem amplification of a pre-binding site for the MSL-complex on the neo-X (5 and 9 tandem copies, respectively). This may increase the affinity of the MSL-complex for such a genomic location and create a more efficient CES. Indeed, we find that CES containing multiple non-overlapping MREs are more strongly bound by the MSL-complex compared to those with a single MRE present (p = 0.038 one-tailed Wilcoxon test, **Supplementary Figure** 2). Thus, a broad spectrum of mutational events has contributed to the evolution of novel CES on the neo-X. Only half of the CES identified on the neo-X required the presence of a pre-site, suggesting that the acquisition of dosage compensation is not necessarily constrained by the fortuitous presence of a sequence that resembles the MSL-recognition motif.

**Heterochromatin formation on the neo-Y and repetitive elements**. The neo-Y, in contrast, is beginning to evolve a heterochromatic appearance [26,27]. Immunostaining of polytene chromosomes demonstrates that the neo-Y is highly enriched for histone modification H3K9me2 (a modification characteristic of heterochromatin) and bound by HP1a (heterochromatin protein marker, recognizing H3K9me2), relative to the neo-X or the rest of the genome (**Figure** 1). We obtained ChIP-seq profiles of H3K9me2 to confirm enrichment of this repressive histone mark on the neo-Y and identify sequence features that are associated with heterochromatin formation. Analyses of the neo-Y genomic sequence is complicated by two characteristics of the neo-Y [24,25,27,34]: On one hand, the neo-Y is highly repetitive, resulting in a fragmented *de novo* genome assembly; on the other hand, unique sequences on the neo-Y are rather similar to their neo-X homologs, and not all read-pairs can be mapped unambiguously to either the neo-X or neo-Y. We thus identified diagnostic SNPs between the neo-X and neo-Y chromosome based on comparisons of male and female genomic libraries, which were used to determine relative enrichment of H3K9me2 at neo-Y versus neo-X gene regions (see Methods for more details). Indeed, we find that the repressive histone mark H3K9me2 is highly enriched at neo-Y genes relative to their neo-X homologs (**Figure** 3A,B), consistent with the polytene chromosome immunostaining results. The initiation of heterochromatin at a specific genomic region and subsequent spreading is less well understood in Drosophila, but is thought to be triggered by the presence of repetitive DNA [11,12]. The neo-Y of *D. miranda* shows a striking enrichment of transposable elements, with about 30-50% of its DNA being derived from repeats [24-27], and the genome assembly of the neo-Y is highly fragmented due to its high repeat content [25]. To assay if transposable elements contribute to heterochromatin formation on the neo-Y, we measured local repeat density around focal genes and their up/down stream regions on the neo-Y relative to the neo-X, by taking advantage of mate-pair relationships of genomic libraries from male *D. miranda*. In particular, we anchored the genomic reads to neo-X/neo-Y diagnostic SNPs, and assayed which fraction of mate-pair reads would map to a repeat library generated for *D. miranda* (see Materials for details). Indeed, we found that neo-Y genes in regions of higher repeat density show elevated H3K9me2 binding levels (**Figure** 3C, linear correlation *P*-value=0.000308). Thus, our data support current models of heterochromatin formation with repetitive elements enabling initiation or spreading of heterochromatin along the neo-Y.

**Interaction of dosage compensation, neo-Y gene decay and heterochromatin formation.** Dosage compensation is thought to evolve in direct response to Y degeneration [35]. Heterochromatin formation, on the other hand, could either be an adaptation to silence maladaptive genes on the neo-Y or genes whose homologs are dosage compensated on the neo-X [36], or it could have deleterious consequences if silencing arises at potentially functional genes [37]. If gene decay on the Y chromosome is driving the evolution of dosage compensation [35,38], neo-X genes with a nonfunctional neo-Y copy should be preferentially bound by the dosage compensation complex. However, spreading of the MSL-complex also implies that neo-X genes with functional neo-Y homologs can become dosage compensated if they reside close to a CES. We divide genes on the neo-Y into non-functional genes if they contain frame-shift mutations or premature stop codons, or if they are deleted from the neo-Y and potentially functional genes if they have an intact open reading frame. Note that the potentially functional genes might nevertheless contain amino-acid substitutions that render them non-functional, or they may contain disabling mutations in their regulatory regions and may not be expressed on the neo-Y. In total, 22% of all annotated neo-sex genes are bound by the MSL-complex on the neo-X (**Supplementary Figure 3**). However, MSL-binding in *D. melanogaster* is more transient than its more broadly distributed chromatin mark H4K16ac [33], and we see a similar pattern in *D. miranda* (**Supplementary Table 1, Supplementary Figure 4**); 44% of all genes on the neo-X are dosage compensated, if compensation is defined by either MSL- and/or H4K16ac enrichment (**Figure** 4A). We find MSL complex binding/H4K16ac enrichment to 46% of the neo-X homologues of neo-Y genes with disrupted ORFs. This value is similar to MSL and/or H4K16ac binding to neo-X genes with intact neo-Y homologues (44% bound on the neo-X; Fisher's exact test *P*-value=0.28 see **Figure** 4A). Further, of neo-X genes whose neo-Y homologs are transcriptionally silent (FPKM<1, see Methods), only 37% are bound by the MSL-complex/enriched for H4K16ac, significantly fewer than neo-X genes with a transcribed neo-Y copy (51%, Fisher's exact test $P$-value=1.6e$^{-12}$, **Figure** 4A). Thus, a large number of genes have become dosage compensated on the neo-X, regardless of whether their neo-Y homologue is functional or not. Moreover, the homologs of genes that are actively transcribed on the neo-Y actually appear more likely to be targeted by the dosage compensation complex on the neo-X, which is contrary to the expectation that dosage compensation has evolved to counterbalance reduced expression of genes that have become silenced on the neo-Y. These patterns of MSL binding are probably due to spreading of the MSL complex in *cis* from CES (see also below). In particular, the recruitment of the MSL-complex to the neo-X by only 68 CES causes dosage compensation of over 1200 transcribed genes. Hence, the acquisition of each CES could have been driven by only a few dosage-sensitive genes on the neo-X (with a nonfunctional neo-Y copy), and most other neo-X genes within each compensated block might have acquired dosage compensation unnecessarily through spreading of the MSL complex. Consistent with spreading passively compensating many neo-X genes, GO enrichment analysis reveals no clear categories of genes as being targeted by the MSL-complex or not (**Supplementary Table** 2). On the other hand, non-functional genes on the neo-Y (those with disrupted ORFs) are more likely to be associated with H3K9me2 (Fisher's exact test, p<0.01, **Figure** 4B). This is consistent with the idea that heterochromatin formation might allow silencing of maladaptive neo-Y genes [36], or silenced genes are free to accumulate nonsense mutations neutrally [37]. However, this association is far from perfect, and many genes with disrupted ORFs (46%) are not silenced by H3K9me2, and many genes with intact ORFs (47%) are targeted by heterochromatin (see **Figure** 4B).

**Gene expression on the evolving sex chromosomes.** Partial degeneration and silencing of neo-Y genes, and incomplete dosage compensation of the neo-X suggest that there may be massive misexpression of neo-sex linked genes in male *D. miranda*. Many genes that are non-functional or silenced on the neo-Y are not yet dosage compensated on the neo-X, while homologs of functional neo-Y genes often reside within dosage compensated blocks on the neo-X. To confirm that MSL binding or enrichment for chromatin marks which are associated with dosage compensation (H4K16ac) result in transcriptional upregulation of neo-X linked genes in *D. miranda*, we compared expression patterns for neo-X genes between males and females in *D. miranda* to their "ancestral" sex-biased expression in *D. pseudoobscura*, where they are autosomal (**Figure** 4C, panel 1). Our genome assembly of the repeat-rich neo-Y is not yet of sufficient quality and contiguity to directly extract genes; instead, we used *de novo* assemblies of the transcriptome to compare transcript abundance between neo-X and neo-Y homologs, and between sexes and species (see Methods). Conditioning on active transcription (based on H3K36me3 enrichment), we find that neo-X genes that are bound by MSL/H4K16ac (or only MSL, **Supplementary Figure** 5) are up-regulated, on average, relative to genes that are not associated with those marks (**Figure** 4C, panel 2). Importantly, however, neo-X genes that lack dosage compensation are not simply transcribed at half the level of genes bound by MSL/H4K16ac (see **Figure** 4C, panel 2). Instead, buffering mechanisms for expression of haploid genes, as generally observed in *Drosophila* [39,40], result in partial compensation of genes that are not targeted by the dosage compensation machinery. Further, many genes are still transcribed from the neo-Y, despite being dosage compensated on the neo-X, or despite harboring frame-shift mutations and stop codons ((**Figure** 4C, panel 3, **Figure** 4D), and there is no statistical association between MSL-binding levels of neo-X genes and down-regulation of their neo-Y homologs (F-statistic test *P*-value=0.73, **Supplementary Figure** 6). In fact, if expression from the neo-Y chromosome is taken into account, many genes appear over-expressed in male *D. miranda* (**Figure** 4C, panel 4). However, it is unclear if the neo-Y copies, which often contain several amino-acid or nonsense mutations [24,25], can functionally substitute for their neo-X homologs. Genes are generally transcribed at a much lower level from the neo-Y relative to the neo-X [25,28] (**Figure** 4C, panel 3, **Figure** 4D), which could in part be caused by changes to its chromatin structure. Consistent with H3K9me2-induced silencing, we find lower expression of H3K9me2 bound neo-Y genes (**Figure** 4E) and a significantly negative correlation between H3K9me2-binding level vs. neo-Y transcript levels (*P*-value<2.2e-16, coefficient=-1.26; **Figure** 4F).

**Chromatin structure evolution.** If epigenetic silencing on the neo-Y evolves in direct response to dosage compensation of neo-X genes, or *vice versa*, we would expect to find the homologs of compensated genes on the neo-X being preferentially targeted by H3K9me2 on the neo-Y. Contrary to this expectation, we detect an overall negative correlation between levels of H4K16ac (or MSL)-binding of neo-X linked loci, and H3K9me2 binding of neo-Y genes (*P*-value=$6.21*10^{-6}$, linear regression coefficient: -0.08, **Figure** 5A, B), and the pattern is more prominent in neo-Y genes bound by H3K9me2 (*P*-value=$4.58*10^{-9}$, coefficient: -0.15). This means that dosage compensated neo-X regions are somewhat less likely to be silenced by heterochromatin on the neo-Y. This negative relationship may in fact reflect different propensities of the ancestral chromosome that formed the neo-sex chromosome to adapt *active* versus *repressive* chromatin configurations. In particular, spreading of the MSL-complex is targeted to actively transcribed regions [41], while heterochromatin is more prone to form in silent, non-transcribed DNA [42]. An ideal outgroup to establish the ancestral chromatin structure of the neo-sex chromosomes would be *D. pseudoobscura*, where this chromosome is still an autosome. In the absence of such data, we used *D. melanogaster* chromatin data and chromatin profiles from *D. miranda* females as a proxy for the ancestral chromatin structure of the neo-sex chromosomes. If we classify *D. miranda* genes according to their principal chromatin types in *D. melanogaster* [43], we observe a general agreement between expression patterns in *D. miranda* and chromatin type (i.e. reduced gene expression in repressive chromatin, and higher gene expression in active chromatin; **Supplementary Figure** 7), suggesting that overall patterns of chromatin structure are conserved between species, and that we can use *D. melanogaster* as a proxy for the ancestral chromatin configuration [43]. We indeed find that genes that are located in active ('yellow') chromatin (**Supplementary Figure** 8) are more likely to have evolved MSL-mediated dosage compensation, while genes in silent ('black') chromatin are more likely to have become heterochromatic on the neo-Y (**Figure** 5B,C). A similar pattern is also found using female *D. miranda* chromatin states for approximating the ancestral chromatin configuration of the neo-sex chromosomes (**Supplementary Figure** 9). While the neo-X is no longer autosomal in this comparison, the chromatin structure (as measured by H4K16ac enrichment) is similar between the X and autosomes in females and differs dramatically in males (**Supplementary Figure** 10), indicating that *D. miranda* females should also provide a good proxy for the ancestral chromatin structure of the neo-sex chromosomes. Chromatin states are overall conserved between *D. melanogaster* and *D. miranda* females, validating our inferences of ancestral chromatin states (**Supplementary Figure** 11).

In addition, homologs of H3K9me2-bound genes are expressed at significantly lower levels in *D. pseudoobscura* compared to homologs of neo-Y genes that are not targeted by H3K9me2 (Wilcoxon Test: W = 746184; *P*-value < 0.01), while genes bound by MSL and/or H4K16ac) on the neo-X tend to have higher expression levels in *D. pseudoobscura* than those not targeted by the dosage compensation machinery (W = 846410.5, p-value < 2.2e-16, **Figure** 5D). This is consistent with heterochromatic regions on the neo-Y being ancestrally less transcriptionally active, while dosage compensation on the neo-X preferentially evolved in transcriptionally active chromosomal segments. Thus, the acquisition of a hyper-transcribed state of the neo-X is accompanied by the acquisition of an inert, heterochromatic chromatin structure on the neo-Y, but neither epigenetic modification appears to directly trigger the other (**Figure** 6).

**Discussion**

*D. miranda* has a unique karyotype, harboring three sex chromosomes of different age: XL is the ancestral X chromosome in the genus Drosophila and >60MY old, XR became X-linked about 15MY ago, is entirely dosage compensated [32] and its former homolog is completely degenerated (i.e. all genes on XR are hemizygous in males [44]). This implies that a former autosome can become completely transformed into a heteromorphic sex chromosome within only 15MY. The much younger neo-sex chromosomes are at an earlier stage of this evolutionary transition, and the neo-Y is only partially degenerated and the neo-X has evolved incomplete dosage compensation. This provides a unique opportunity to study the evolutionary processes driving the differentiation of sex chromosomes, and here we investigate how changes to the DNA sequence result in novel epigenetic modifications of the diverging neo-sex chromosomes that affect levels of transcription of neo-sex linked genes. Recruitment of the dosage compensation complex to the neo-X requires the acquisition of a 21-bp consensus motif, and we uncover diverse mutational paths that have led to the evolution of novel CES on the neo-X. This highlights how evolution can follow predictable genetic trajectories by repeatedly acquiring the same 21-bp consensus motif for recruitment of the dosage compensation complex, yet utilizing a diverse array of random mutational changes to attain the same phenotypic outcome. We further show that heterochromatin formation is triggered by an accumulation of repetitive DNA on the neo-Y, and silences adjacent genes.

Surprisingly, we find little correspondence between Y degeneration and dosage compensation in *D. miranda*. Many non-functional neo-Y genes are not dosage compensated on the neo-X while many potentially functional neo-Y genes reside within dosage compensated blocks. Spreading of the MSL-complex implies that the acquisition of dozens of CES can result in dosage compensation of 100s of genes along the neo-X, many with functional and expressed homologs on the neo-Y. Patterns of gene expression confirm that many neo-sex genes are either over- and under-expressed, i.e. there is rampant suboptimal transcription in male *D. miranda*. Dosage compensation of functional genes and transcription of pseudogenes from the neo-Y may in fact select for adaptive down-regulation of those genes from the neo-Y, i.e. the degeneration of genes on the neo-Y that are dosage compensated on the neo-X would be selectively favored. Thus, once an evolving X chromosome acquires dosage compensation mechanisms that operate through large-scale modifications to its chromatin structure, such as in Drosophila, the entire evolutionary dynamics of sex chromosome evolution will change. While Y degeneration at the initial stages of sex chromosome evolution is a deleterious process with negative consequences to fitness, degeneration of neo-Y genes whose homologs are dosage compensated on the neo-X should restore optimal levels of gene expression, and thus improve fitness. This will result in complex patterns of Y degeneration over evolutionary time, and confounds comparisons of sex chromosome evolution in taxa with different modes of dosage compensation [45].

While heterochromatin formation on the neo-Y occurs simultaneously with dosage compensation on the neo-X in *D. miranda*, the genomic regions that are being targeted by MSL-dependent dosage compensation on the neo-X and heterochromatin on the neo-Y show little overlap. This may reflect different propensities of the ancestral chromosome that formed the neo-sex chromosome to evolve active versus repressive chromatin configurations (**Figure** 6). In particular, the MSL-complex spreads along the X chromosome by targeting actively transcribed regions, and spreading should be more efficient in chromosomal neighborhoods that display higher levels of genetic activity. In contrast, active transcription suppresses the spreading of heterochromatin, and heterochromatin is more likely to form and propagate in genetically inert regions. We show that chromosomal neighborhoods of the neo-sex chromosome with ancestrally higher levels of expression and that are classified as active chromatin are more likely to be targeted by the MSL-complex; in contrast, ancestrally silent chromatin with reduced gene expression is more likely to have adopted a heterochromatic appearance on the neo-Y. Thus, the antagonizing effects of active transcription and associated differences in chromatin structure can help to explain the evolution of epigenetic modifications on diverging sex chromosomes.

The epigenome of sex chromosomes is very different in mammals compared to Drosophila. For one, the Y chromosome in mammals is less heterochromatic than in Drosophila [46]. The human Y contains one large heterochromatic block [47], but most genes appear to reside within the euchromatic part of the chromosome, and the macaque Y chromosome contains almost no heterochromatin [48]. Also, dosage compensation works in opposite directions in mammals and flies, with one of the two X chromosomes being inactivated in female mammals [35]. X inactivation in mammals is initiated in early embryogenesis by *Xist* RNA that localizes to the inactive X chromosome. *Xist* induces X-chromosome inactivation (XCI) by spreading in *cis* across the future inactive X-chromosome [49], recruiting a polycomb repressive complex [50], and forming a transcriptionally silent nuclear compartment enriched for repressive chromatin modifications including H3K27me3 [50]. XCI in mammals is initiated from a single region on the X (the X inactivation center, the genomic location from which *Xist* is being transcribed), while Drosophila contains 100s of CES along its X. Both the repressive chromatin modification in mammals and the active chromatin in Drosophila spread across the chromosome in a sequence-independent manner. During initiation of XCI, *Xist* transfers to distal regions across the X-chromosome by exploiting the three-dimensional conformation of the X-chromosome; i.e. *Xist* coats the X-chromosome by searching in three dimensions, modifying chromosome structure, and spreading to newly accessible locations [51]. The MSL-complex of Drosophila spreads along the X chromosome by recognizing features of actively transcribed genes (i.e. H3K36me3 modification), but it is not known if MSL spreads linearly along the X chromosome, or in three dimensions as well. During the maintenance of XCI in mammals, *Xist* binds broadly across the X-chromosome [51] while the MSL-complex in flies is highly enriched in actively transcribed genes [10]. While we find little correspondence between whether a gene is dosage compensated on the neo-X and whether its neo-Y homolog is functional in *D. miranda*, X inactivation in humans is primarily driven by gene loss on the Y, and X-inactivation status can successfully classify 90% of X-linked genes into those with functional or nonfunctional Y homologs [52].

To conclude, our study highlights both the potential and the limitations of adaptation. On one hand, we show that the neo-X has rapidly evolved dosage compensation, and makes use of different mutations to acquire MSL-binding motifs. This illustrates how evolution can repeatedly attain the same phenotypic outcome, yet utilizing diverse underlying mutational paths, and demonstrates how random *de novo* mutations and natural selection can quickly respond to fitness costs resulting from gene decay on the neo-Y by co-opting the existing dosage compensation machinery. On the other hand, the peculiar mechanistic property of the MSL-complex to spread along the chromosome results in suboptimal patterns of dosage compensation on the neo-X, causing compensation of many functional neo-Y genes. This in turn sets the stage for adaptive Y-degeneration to restore optimal expression levels of dosage compensated neo-sex linked genes. Thus, our study reveals a dynamic interplay between Y degeneration and dosage compensation, and shows how epigenetic modifications drive the evolution of silent and hyper-transcribed chromatin on evolving sex chromosomes, though neither directly triggers the other.

**Materials and Methods**

**SNP calling and neo-Y annotation.** We sequenced single individuals of both sexes from an inbred *D. miranda* strain (MSH22) at ~90 fold coverage for each sex. The genome assembly and annotation has been greatly improved relative to the earlier version presented in [25] (N50 length: 1,029kb vs. 23.7kb), because of the increased sequencing coverage and inclusion of Illumina libraries with different insert sizes and 454 data [32]**.** We aligned the genomic reads of male and female against this improved version of *D. miranda* chromosome sequences using **bowtie2** [53] using the 'sensitive-local' parameter set and taking the read orientation and library insert size into consideration, and then screened the alignments by their mapping qualities (Q>20, where 'Q' is the mapping quality determined by bowtie2 and Q>20 means a certain alignment has less than 1% chance to be spurious). Following the standard **GATK pipeline** [54], PCR duplicate reads were removed and reads were realigned before calling variants with **UnifiedGenotyper**. We discarded SNPs/indels with low qualities (Quality<30) or coverage (Depth<5) or showing unusual strand-biases or clustering patterns. Since we have sequenced single individuals of an inbred *D. miranda* strain (3 libraries per sex), male-specific variants linked to the neo-sex chromosomes should likely represent neo-Y specific mutations. We have identified a total of 380,684 such mutations (putative neo-Y specific mutations), translating to an average divergence level of 1.8 sites per 100bp between the neo-X/Y. We estimated the false positive discovery rate to be about 2% to 4%, based on the numbers of male-specific variants on autosomes. The putative neo-Y specific mutations were introduced into the neo-X chromosome sequence to build a reference-based neo-Y chromosome assembly. There are a total of 169,046 neo-X/Y divergence sites identified from 2496 neo-sex linked genes (92.4% of all annotated genes) with an average of 67 divergent sites per gene. This provides diagnostic sites dense enough for our further discrimination between neo-X and neo-Y Chip-seq /RNA-seq reads. We then used predicted neo-X protein sequences to annotate the reconstructed sequence of the neo-Y, and any genes containing premature stop codons or frameshift mutations were characterized as neo-Y genes with disrupted ORFs. We inferred genes deleted from the neo-Y by comparing the mapping coverage between sexes (**Supplementary Figure 12**) and conditioned on a lack of male-specific variants in such genes; neo-Y deletion genes are defined as those showing the same distribution of mapping coverage between sexes as X-linked genes. Note that most of our analysis of neo-Y chromosome features is done relative to the neo-X focusing on these neo-X/Y divergent sites; the few genes / gene regions that lack diagnostic SNPs between the neo-sex chromosomes should not greatly bias our analysis (6.6% of all genes).

**Evolutionary analysis**. To infer the molecular evolution and conservation of CES, we used the software package **Mercator** [55] to generate whole-genome alignments between *D. miranda*, *D. pseudoobscura* and the more distant outgroup *D. affinis*, for comparison along the neo-X, and *D. melanogaster* for contrasts of CES on XL. We used **FIMO** [56], part of the **MEME** [57] suite, to identify genomic regions showing homology to the MRE motif and extracted from the whole-genome alignment the highest scoring motif within 500bp of each CES summit on the neo-X chromosome. We manually examined each alignment to infer the mutational path by which the motif arose in *D. miranda*.

**Polytene chromosome immunostaining & ChIP-seq**. Polytene chromosomes were isolated from male 3[rd] instar larvae and processed for immunostaining as described [8]. Chromatin immunoprecipitation from sexed male and female third instar larvae were prepared as described [8]. The following antibodies against histone modifications were used for ChIP-seq experiments: (1) H3K9me2 (Abcam ab1220; 3 μl/IP); (2) anti-H3K27me3 (Abcam ab6002; 5 μl/IP); (3) anti-H3K36me3 (Abcam ab9050; 3 μl/IP) and (4) anti-H4K16ac (Millipore 07-329; 5 μl/IP). Immunoprecipitated and input DNAs were purified and processed according to the standard paired-end Solexa library preparation protocol. Paired-end 100-bp DNA sequencing was performed on the Illumina Genome Analyzer located at UC Berkeley Vincent J. Coates Genomic Sequencing Facility. The following data set were used from [32] (1) anti-H3K36me3 male third instar larvae; (2) anti-H4K16ac male third instar larvae; (3) MSL3-TAP mixed-sex larvae, accession numbers SRS402820 and SRS40282.

**ChIP-seq analysis.** We aligned the ChIP-seq and input control reads against the reference genome using **bowtie2** and then separated them into neo-X or neo-Y linked reads using male-specific variants. Only reads containing diagnostic variants that allow us to distinguish between the neo-X and the neo-Y allele are used for this analysis, and we only kept reads that have a mapping quality of >30 (such reads have a <0.001 chance to be misidentified as a result of misalignment) and we further require each diagnostic site to have at least three reads for both neo-X and neo-Y alleles to be considered (see **Supplementary Table** 3). Removing regions with no input signal, Log2 mapping coverage ratio of ChIP vs. control was investigated along the gene body, including 3kb of up- and downstream regions, to reflect the binding intensities of certain chromatin markers. The distributions of binding intensities usually show a distinctive bimodal pattern on sex or neo-sex chromosomes compared to autosomes; thus we defined the bound/unbound genes for each chromatin markers at the values where the two peaks of distribution separated out (**Supplementary Figure 13**). We also extracted the chromatin state 'color' information for all the *D. melanogaster* genes from [43]. To associate such information with *D. miranda* genes, we used ortholog information between *D. pseudoobscura* and *D. melanogaster* retrieved from FlyBase.

**Repeat analysis.** We generated a consensus *D. miranda* repeat library with **RepeatModeler** and **RepeatMasker** (http://www.repeatmasker.org), using both the latest *D. miranda* genome assembly (from females) [32] and a previous *de novo* assembly of the neo-Y [25]. We mapped reads from a genomic library of *D. miranda* males (less than 1kb insert size) against neo-sex linked genes and their flanking regions using **bowtie2** with single-end reads mapping mode, and 'sensitive-local' option, and assigned linkage of the reads to the neo-X/Y according to male-specific diagnostic SNPs (**Supplementary Figure 14**). We then mapped the other mate pair of the neo-X or neo-Y specific read against the repeat consensus library, to estimate local repeat density at neo-X versus neo-Y focal genes. The mapping was done using **bowtie2** with single-end reads mapping mode, and 'very-sensitive-local' parameter set.

**Gene expression analysis.** Our genome assembly of the highly repeat-rich neo-Y is not yet of sufficient quality and contiguity to directly extract genes from the *de novo* assembly. Most of our analysis studying the chromatin structure of the neo-Y, or its genomic composition (i.e. analysis of the ChIP-seq data, or TE enrichment on the neo-Y) was done relative to the neo-X. For this analysis, reconstructing the neo-Y sequences as outlined above by introducing male-specific variants was appropriate. To study the transcriptome, we wanted to compare expression levels from the neo-sex chromosome to their ancestral expression levels in *D. pseudoobscura* and contrast expression in males vs. females, to test for an up-regulation of dosage compensated neo-X genes, and down-regulation of neo-Y transcripts. For this analysis, we required absolute expression levels of neo-sex transcripts, to compare across sexes and species, and we generated *de novo* transcriptome assemblies for *D. miranda* and *D. pseudoobscura*, using **trinity** [58]. The pipeline for the assembly of the neo-sex transcriptomes will be described in more detail (Kaiser and Bachtrog, in prep.); briefly, neo-X transcripts were re-constructed using a **trinity** transcriptome assembly from females, and neo-Y transcripts were re-constructed using a **trinity** transcriptome assembly from males, which was modified to contain all neo-Y-specific variants; this procedure was necessary to resolve chimeric neo-X/Y transcripts produced by **trinity**. In particular, sections of neo-Y transcripts were kept for the final assembly only if they contained at least one neo-X/neo-Y distinguishing variant, and if they were fully supported by RNA-Seq reads; and genes inferred to be deleted from the neo-Y were excluded from the neo-Y assembly. The neo-sex transcriptome has been submitted to GenBank, accession numbers GALP00000000. To calculate transcript abundance, neo-X and neo-Y RNA-Seq reads from male and female larvae were mapped against the neo-sex chromosomal transcripts using **Mosaik** (http://bioinformatics.bc.edu/marthlab/wiki/index.php/Software), allowing for zero mismatches, i.e. reads were exclusively assigned to their respective neo-sex chromosomes of origin, whenever there was a SNP or indel present. **eXpress** [59] probabilistically assigns all reads to alleles (including reads mapping to both the neo-X and neo-Y) and was used to calculate transcript abundance (FPKM: Fragments Per Kilobase of transcript per Million mapped reads) for the neo-X in *D. miranda*, separate from any neo-Y expression, and *vice versa*; similarly, **eXpress** was used to calculate transcript abundance in *D. pseudoobscura*. We defined a neo-Y gene to be actively transcribing if its FPKM value is higher than 1, which is derived as a cut-off from comparing FPKM distributions of genes vs. intergenic regions (**Supplementary Figure** 15). FPKM values for each gene are given in **Data S1**.

**ACKNOWLEDGMENTS**

AAA & AAG thank Dr. M.I. Kuroda for support, in whose laboratory the ChIP and immunostaining experiments were performed. The ChIP-seq data has been deposited in NCBI Short Reads Archive under the accession number SRR899838, and RNA-seq data has been deposited under the accession number SRR899847 & SRR899848. The transcriptome shotgun assembly project has been deposited at DDBJ/EMBL/GenBank under the accession GALP00000000. The version described in this paper is the first version, GALP01000000.

**Figures**

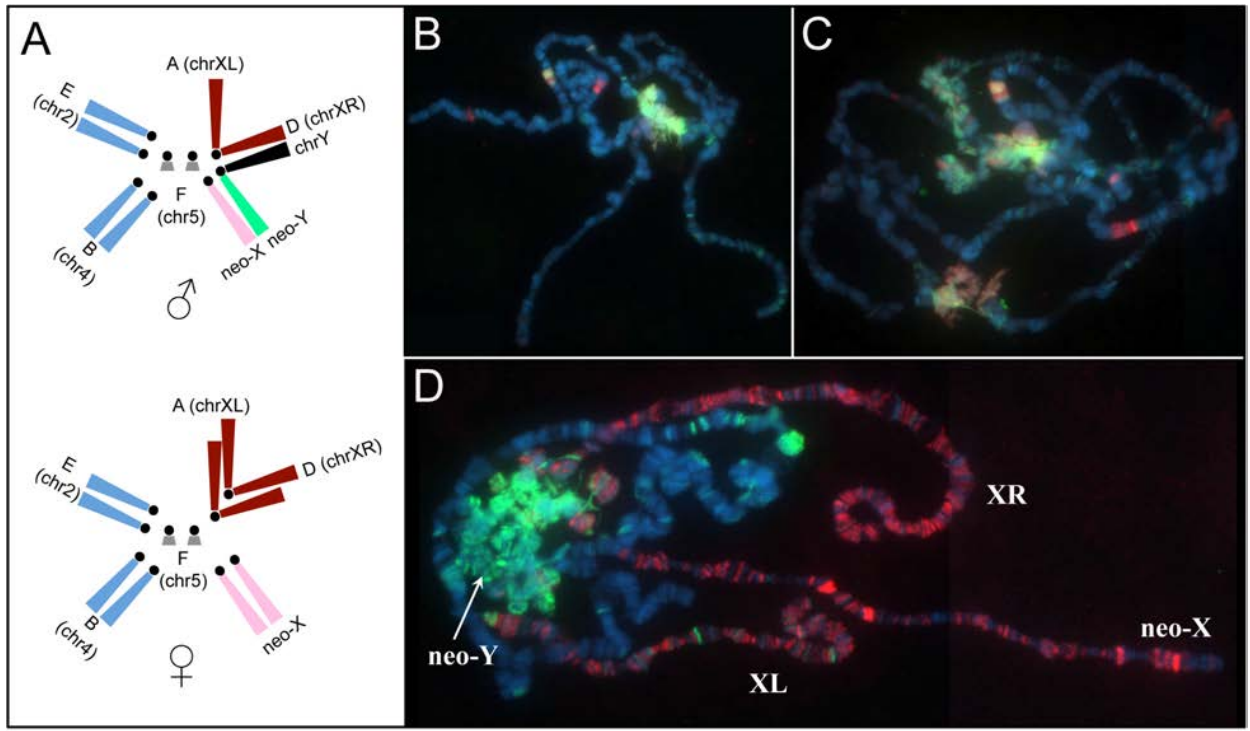

**Figure 1 Dosage compensation of the neo-X, and heterochromatin formation on the neo-Y of *D. miranda*.** (**A**.) Schematic karyotype of *D. miranda*. *Drosophila* chromosomes are labeled as 'Muller element' from A to F. In *D. miranda*, two fusions between element A (ancient X) and D, and the Y chromosome and element C created sex chromosomes of different ages. Element D became chrXR about ~10-15 million years ago and element C became the neo-X and neo-Y chromosome about ~1-1.5 million years ago. (**B.-C.**) Polytene chromosomes stained for H3K9me2 (green) and HP1a (red) in (**B.**) female *D. miranda* and (**C**.) male *D. miranda*. (**D**.) Co-immunolocalization of MSL3-TAP (red) and H3K9me2 (green) in transgenic male *D. miranda* expressing TAP-tagged MSL3. The neo-Y is becoming heterochromatic, as shown by prominent H3K9me2 and HP1 binding, while all three X-chromosome arms are acquiring dosage compensation in *D. miranda* males.

**Figure 2 Acquisition of CES on the *D. miranda* neo-X chromosome. (A.)** The MSL-recognition element (MRE) identified in *D. miranda* [32]. **(B.)** Number of occurrences of the different mutational events identified to create a MRE on the neo-X. The "Undecipherable" category refers to CES where no MRE was detected on the neo-X ('No motif hit'; 6 CES), or where *D. miranda* had an equally scoring ('Presite'; 12 CES) or lower scoring motif than the outgroup species ('Lower scoring motif'; 7 CES). This suggests that some of the CES may be false positives (i.e. they are highly bound by the MSL-complex through spreading rather than through MRE-mediated targeting) or that secondary mutations in adjacent regions occurred to enable efficient recruitment of the MSL complex to suboptimal MRE's on the neo-X. **(C.)** Examples of different mutational events identified on the neo-X to create a novel MRE. Multiple species alignments are shown for dmir: *D. miranda*, dpse: *D. pseudoobscura*, daff: *D. affinis*, and the MRE element is highlighted in grey.

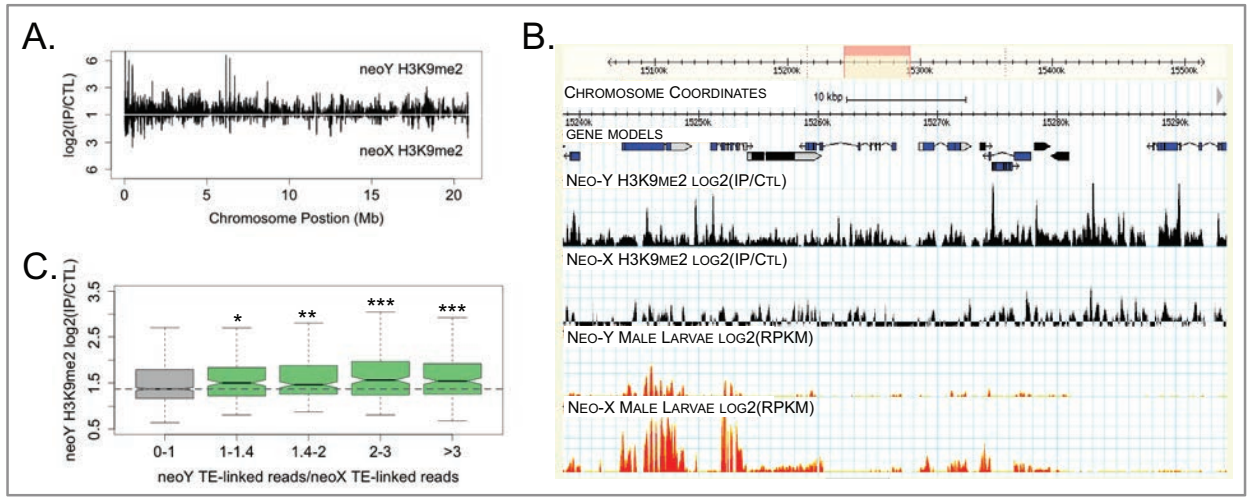

**Figure 3 Heterochromatin formation on the neo-Y of *D. miranda*.** (**A.**) Enrichment profile of H3K9me2 on the *D. miranda* neo-sex chromosomes. Intensity ratios are plotted for H3K9me2 (y-axis) relative to chromosomal position (x-axis), for protein-coding genes and their flanking regions along the neo-sex chromosomes. (**B**.) Genome Browser screen capture of a 50kb region on the neo-sex chromosomes showing intensity ratios for histone marks (H3K9me2, in black) and read coverage depth for RNA-seq data (in red) for the neo-Y and neo-X chromosomes in male third instar larvae. Gene models for potentially functional neo-Y genes are in blue, and for non-functional neo-Y genes in black. (**C.**) TE accumulation on the neo-Y relative to the neo-X, vs. H3K9me2 binding along the neo-Y. The ratios of neo-Y repeat-linked read numbers vs. neo-X repeat-linked reads were pooled into 4 bins of equal size, as a reflection of the degree of neo-Y specific repeat accumulation. The boxplots show the neo-Y specific H3K9me2 binding ratios within each bin, and genes without neo-Y specific repeat enrichments show a significantly lower H3K9me2 binding (Wilcoxon one tailed test: *P*-value<0.05) than others. The number of asterisks reflects the significance level. '*': *P*-value<0.05, '**': *P*-value<0.01, '***': *P*-value<0.0001.

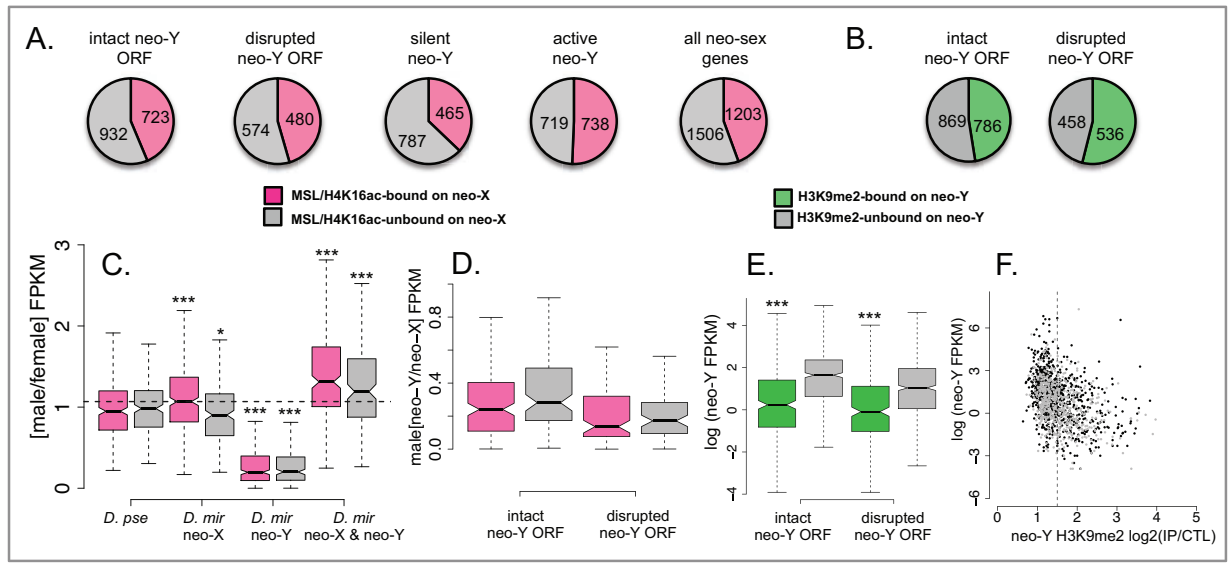

**Figure 4 Dosage compensation and gene silencing.** Genes that are targeted by the MSL complex or enriched for H4K16ac on the neo-X are shown in pink, and genes that are neither bound by MSL nor H4K16ac are shown in grey. (**A**.) The proportion of MSL-bound/H4K16ac enriched genes does not differ between neo-X genes whose neo-Y homologues are potentially functional (intact neo-Y ORF) versus those whose neo-Y homologues are non-functional (disrupted neo-Y ORF). Genes that are transcriptionally silent on the neo-Y (silent neo-Y, FPKM < 1) are less likely to be dosage compensated on the neo-X, while actively transcribed neo-Y genes (active neo-Y, FPKM > 1) are more often dosage compensated. (**B**.) Pseudogenes on the neo-Y (disrupted neo-Y ORF) are significantly more likely to be targeted by H3K9me2 than potentially functional neo-Y genes (Fisher's exact test, p<0.01). (**C**.) Upregulation of gene expression by the dosage compensation complex on the neo-X. Shown is the expression of Muller C genes in males vs. females (M/F), and genes are divided into those bound by the MSL complex and/or H4K16ac-marked on the neo-X in *D. miranda* (pink) versus those not targeted by the dosage compensation machinery on the neo-X (grey). Only transcripts with FPKM > 2 are included. *D. pse* (panel 1): M/F expression of Muller C genes in *D. pseudoobscura*. *D. mir* neo-X (panel 2): Expression of the neo-X allele in males vs. females. M/F expression is significantly higher for genes targeted by the dosage compensation complex compared to neo-X genes that are not targeted (Wilcoxon Test: W = 64915, p < 10⁻⁴), whereas the M/F ratio in *D. pseudoobscura* is indistinguishable between homologs of bound and unbound genes (Wilcoxon Test: W = 51929, NS). Haploid output of dosage compensated neo-X genes is slightly higher than diploid expression of *D. pseudoobscura* homologues (Wilcoxon Test: W = 149307, p < 0.01) (panels 1 & 2) whereas haploid output of unbound neo-X genes is not increased to the same extent, i.e. it is significantly lower compared to diploid expression in *D. pseudoobscura*

(Wilcoxon test: W 18272, p < 0.05) (panels 1 and 2). *D. mir* neo-Y (panel 3): Expression of the neo-Y allele in males vs. the neo-X in females. Neo-Y expression is significantly reduced compared to neo-X expression (MSL/H4K16me3 genes: Wilcoxon Test: W = 17481, p < $10^{-15}$; genes not targeted by MSL/H4K16me3: Wilcoxon Test: W = 3006, p < $10^{-15}$) (panels 2 &3). *D. mir* neo-X & neo-Y (panel 4): Adding up the FPKM-values of neo-X and neo-Y linked genes leads to an estimate of the overall output from the neo-sex chromosomes. Combined neo-sex expression is significantly higher than autosomal expression of homologs in *D. pseudoobscura* (MSL/H4K16me3 genes: Wilcoxon Test: W = 123412, p-value < $10^{-15}$; genes not targeted by MSL/H4K16me3: Wilcoxon Test: W = 17455, p < $10^{-5}$) (panels 1 & 4). (**D**.) As in (C), neo-Y genes whose homologues are dosage compensated on the neo-X are shown in pink, and neo-Y genes with un-compensated neo-X homologues are shown in grey. Neo-Y/neo-X transcript levels are indistinguishable comparing genes with intact neo-Y ORF vs. disrupted neo-Y ORF (Wilcoxon Test: W = 13588, NS, and W = 7096, NS), suggesting that down-regulation of the neo-Y occurs independently of dosage compensation on the neo-X. However, absolute expression of non-functional neo-Y genes is lower compared to that of functional neo-Y genes (Wilcoxon Test: W = 52980, p < $10^{-4}$). (**E**.) H3K9me2-bound neo-Y genes (shown in green) are expressed at significantly lower levels than genes not targeted by H3K9me2 on the neo-Y (shown in grey) (Wilcoxon Test: W = 57006; p < $10^{-15}$ (functional genes) and W = 20662; p < $10^{-15}$ (pseudogenes); all FPKM-values are included). (**F**.) Downregulation of neo-Y genes that are targeted by H3K9me2. Potentially functional neo-Y genes are shown in black, pseudogenes in grey; the vertical line indicates the cut-off value for H3K9me2-bound versus unbound genes.

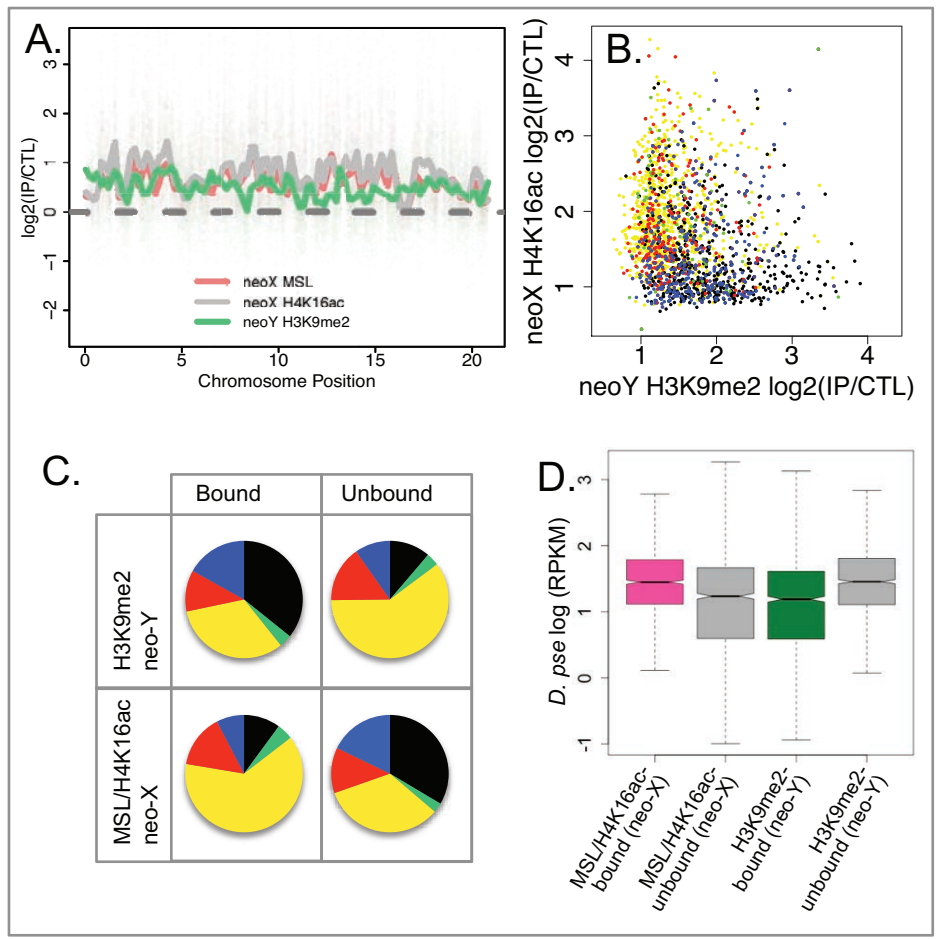

**Figure 5 Heterochromatin formation and dosage compensation** (**A**.) Sliding window enrichment profile of H3K9me2-enrichment along neo-Y genes, and H4K16ac and MSL-binding along their neo-X homologs. (**B**.) H4K16ac-enrichment of neo-X genes versus H3K9me2-enrichment at their neo-Y homologs. Genes are color coded according to their chromatin state in *D. melanogaster* [43], with yellow and red genes corresponding to actively transcribed genes, and black, green and blue genes corresponding to silenced genes. (**C.**) MSL/H4K16ac-bound/unbound neo-X genes and H3K9me2-bound/unbound neo-Y genes vs. principle chromatin types in *D. melanogaster*. The color-coded chromatin types of *D. miranda* bound/unbound genes were inferred from the chromatin type definition of their *D. melanogaster* orthologs (from [43]). Genes within 'yellow' chromatin are more likely to be targeted by the dosage compensation complex on the neo-X. Genes within 'black' chromatin are more likely to be silenced by H3K9me2 on the neo-Y. Genes within active 'red' chromatin show no significant difference regarding their dosage compensation states on the neo-X, which is consistent with the lack of H3K36me3 chromatin mark in red chromatin, and the dosage compensation complex targeting genes with such a mark. (**D**.) Expression levels of genes in *D. pseudoobscura* whose homologs in *D. miranda* are bound/unbound by MSL/H4K16ac on the neo-X or bound/unbound by H3K9me2 on the neo-Y; *D. pseudoobscura* expression levels can be used as a proxy for ancestral expression of neo-sex linked genes.

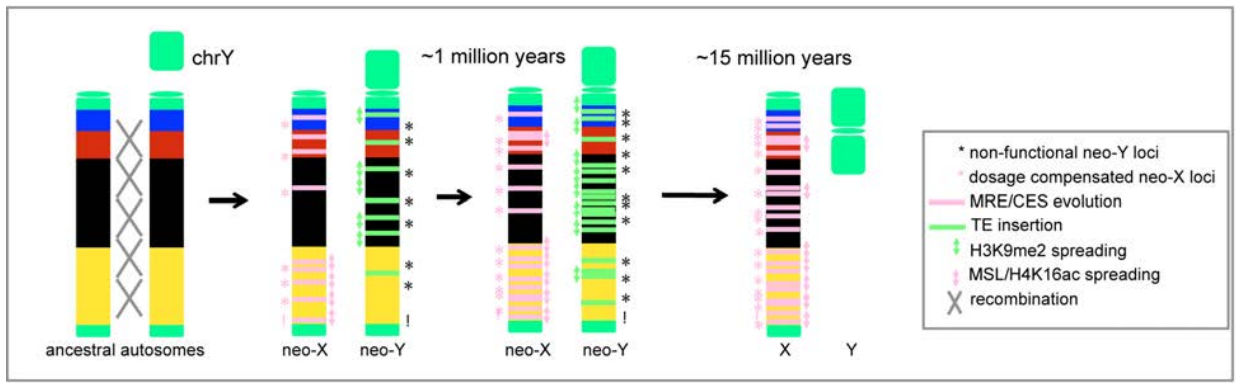

**Figure 6 Model of chromatin changes at evolving neo-sex chromosomes**. The process of heterochromatin formation of the neo-Y chromosome appears to be initiated from repressive (black) chromatin regions and dosage compensation on the neo-X preliminary evolves from active (yellow) chromatin.

**Supplementary Figures and Tables**

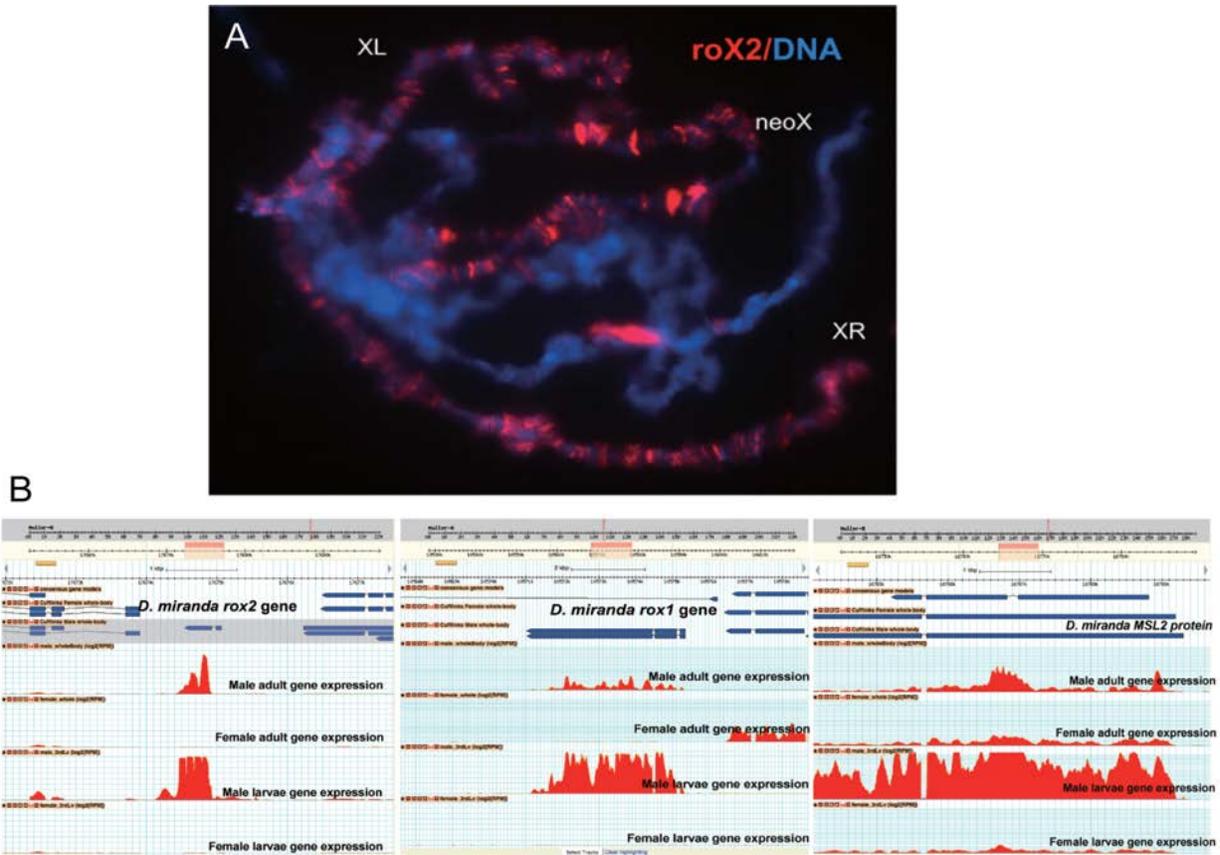

**Supplementary Figure 1 Male-specific targeting and expression of the MSL-complex in *D. miranda*. A.**
*roX-2* RNA-FISH of *D. miranda* male salivary glands. We cloned the *roX2* gene and performed RNA-FISH,
using a similar protocol as described in [60]. B. Male-specific expression of *MSL-2*, *roX-1* and *rox-2*.

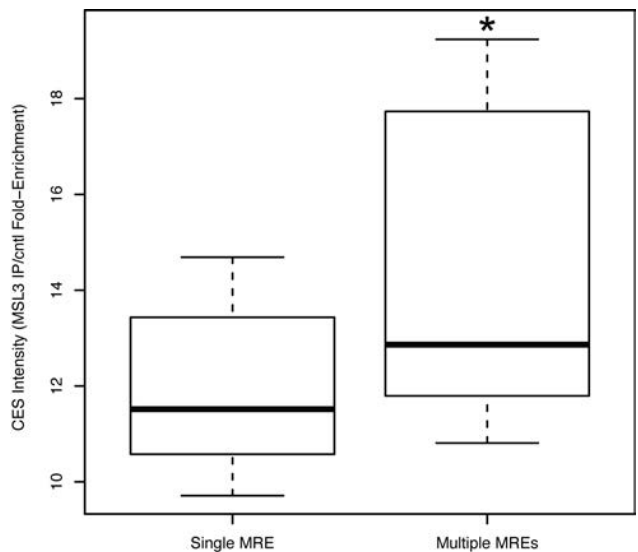

**Supplementary Figure 2.** Intensity of MSL-binding versus number of MRE motifs found at CES. CES with
multiple MRE's show significantly more MSL-binding, than CES with single MRE's (one-tailed Wilcoxon
test p = 0.038).

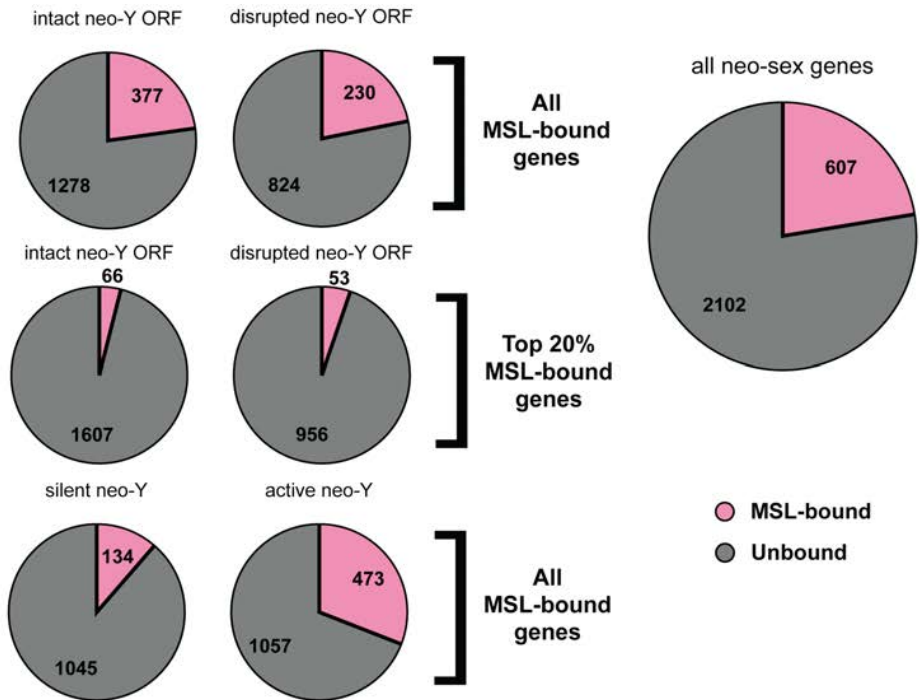

**Supplementary Figure 3. Dosage compensation and neo-Y degeneration.** Genes that are targeted by the MSL complex on the neo-X are shown in pink, and genes that are not bound by MSL are shown in grey. **(A**.) The proportion of MSL-bound enriched genes does not differ between neo-X genes whose neo-Y homologues are potentially functional (intact neo-Y ORF) versus those whose neo-Y homologues are non-functional (disrupted neo-Y ORF). Genes that are transcriptionally silent on the neo-Y are less likely to be dosage compensated on the neo-X, while actively transcribed neo-Y genes are more often dosage compensated.

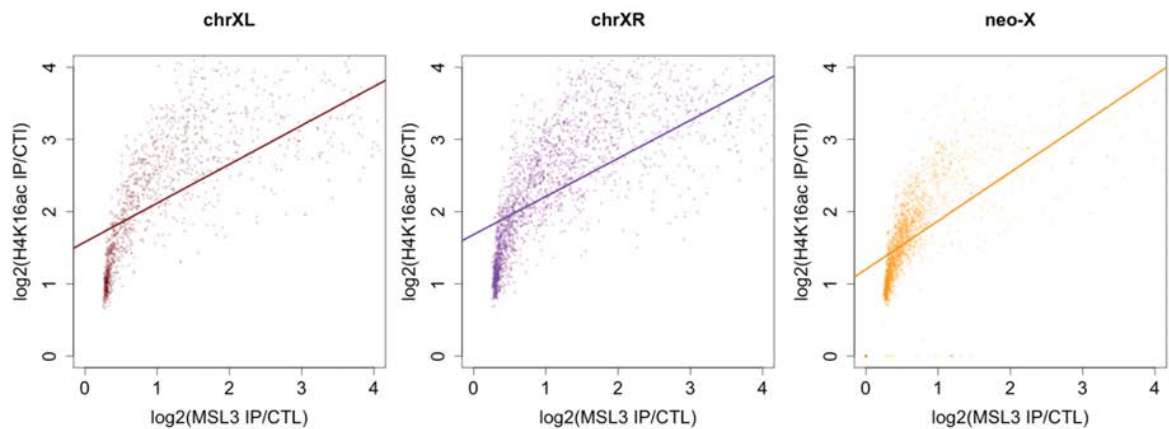

**Supplementary Figure 4. MSL3 enrichment level is significantly correlated with that of H4K16ac.** We show here dot plots of log2 read depth ratios of ChIP-seq vs. input control along the gene body for MSL3 and H4K16ac chromatin marker on different X chromosomes, which significantly correlate with each other (R-square=0.47-0.49, p-value<2.2e-16).

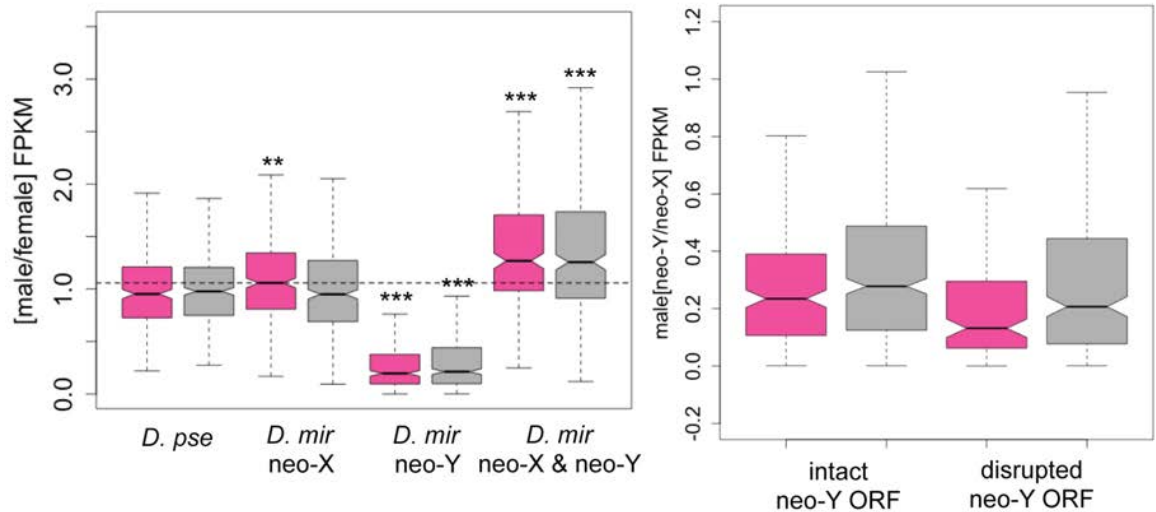

**Supplementary Figure 5** Pattern of dosage compensated genes that are defined by MSL binding only. We have observed the similar pattern comparing as Figure 4C and Figure 4D if we define dosage compensated genes on the neo-X only by significant MSL binding.

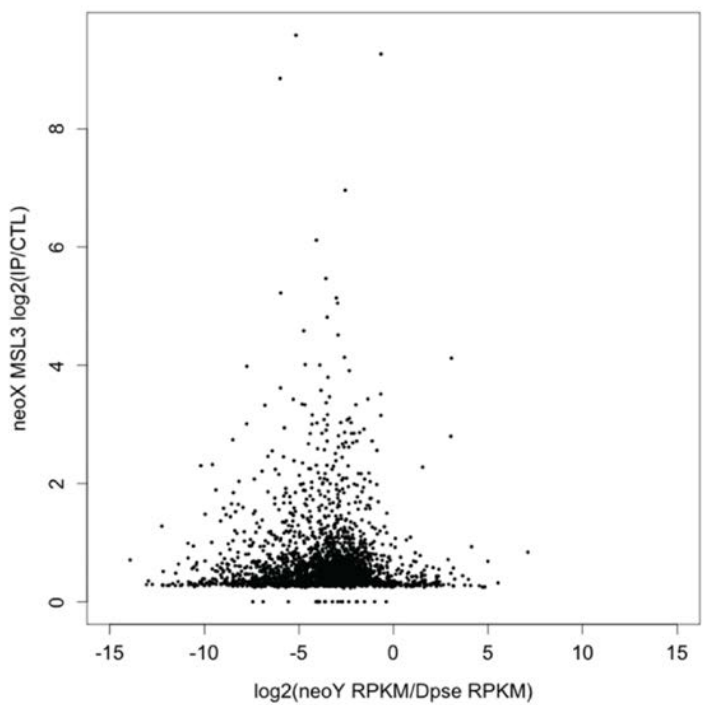

**Supplementary Figure 6. No correlation of neo-Y down-regulation vs. neo-X dosage compensation.** The x-axis shows reduction of neo-Y expression level measured as the log2 ratio of neo-Y gene specific FPKM values vs. those of *D. pseudoobscura* orthologs against the MSL-binding enrichment ratio of their corresponding neo-X genes.

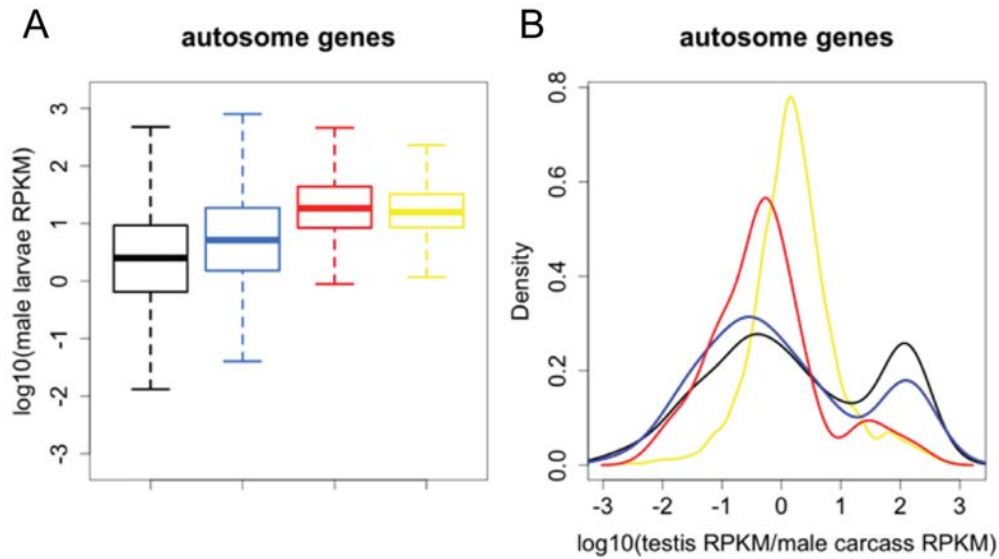

**Supplementary Figure 7. Gene expression patterns for autosomal genes in *D. miranda*, classified by their different chromatin types defined in *D. melanogaster*. A.** We find characteristic *D. miranda* gene expression patterns of each chromatin type that is similar to that of *D. melanogaster* (i.e. reduced gene expression in repressive 'black' or 'blue' chromatin, and higher gene expression in active 'red' or 'yellow' chromatin). B. Genes in black and blue chromatin are more tissue-specific (measured by testis-specificity in *D. miranda*), consistent with their patterns of tissue-specific expression in *D. melangoaster*. These consistent expression patterns between species suggest that we can approximate the *D. miranda* ancestral chromatin types by their *D. melanogaster* orthologs.

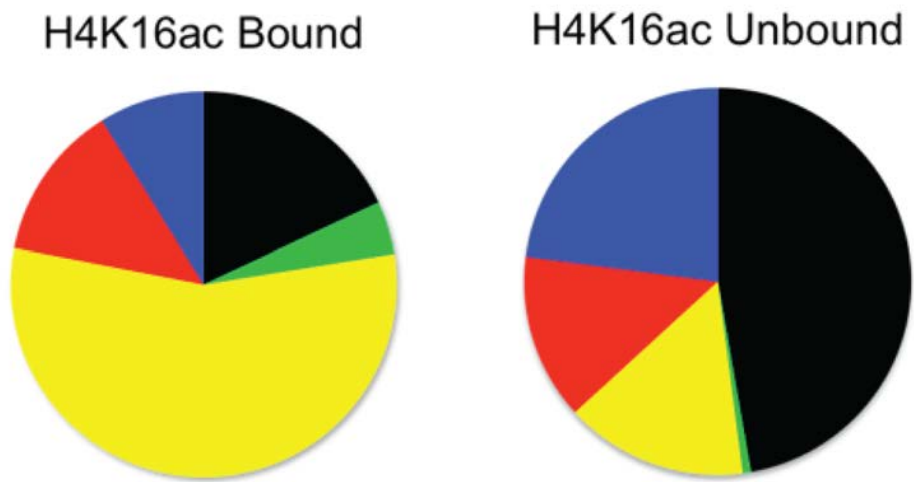

**Supplementary Figure 8. Ancestral chromatin states of H4K16ac bound/unbound genes on chrXR**. ChrXR (the Muller D element) is another young X chromosome that originated around 10 million years ago in an ancestor of *D. miranda* and *D. pseudoobscura*, and has evolved full dosage compensation. Dosage compensated genes on XR (defined as those bound by H4K16ac chromatin marks) are enriched for genes within an active chromatin state ('yellow' chromatin) of *D. melanogaster*.

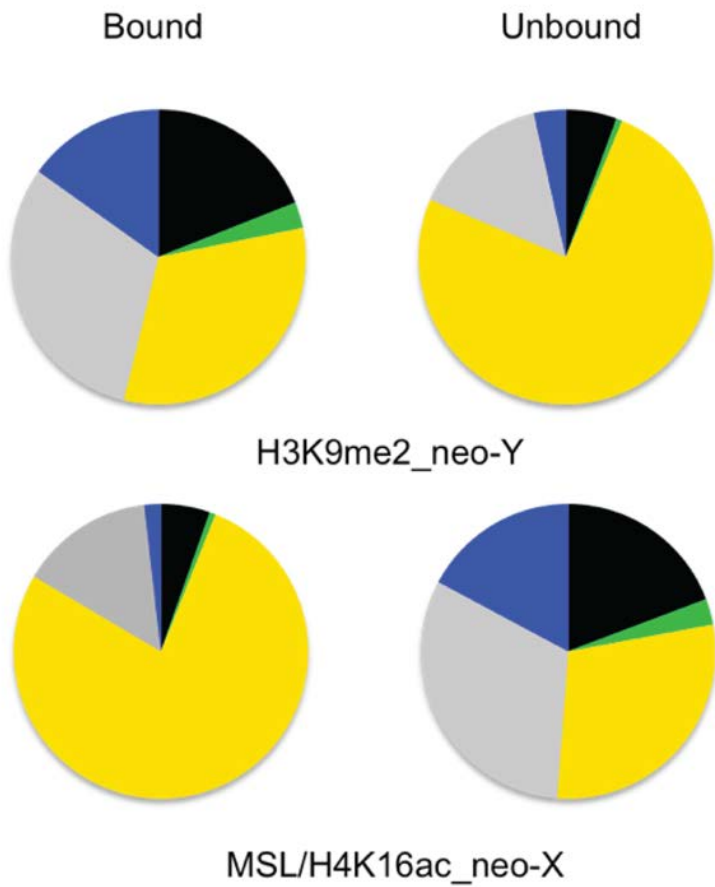

**Supplementary Figure 9. H4K16ac-bound/unbound neo-X genes and H3K9me2-bound/unbound neo-Y genes vs. chromatin states of female *D. miranda*.** We approximate the ancestral chromatin states by ChIP-seq data of female *D. miranda* larvae: blue genes were defined by their characteristic H3K27me3 bound state, green genes by H3K9me2, yellow genes by H3K36me3 and high expression level (FPKM>2), while black genes are not bound by any studied histone markers and show a low expression level (FPKM<2). We grouped the rest of the genes into an unclassified category as grey genes.

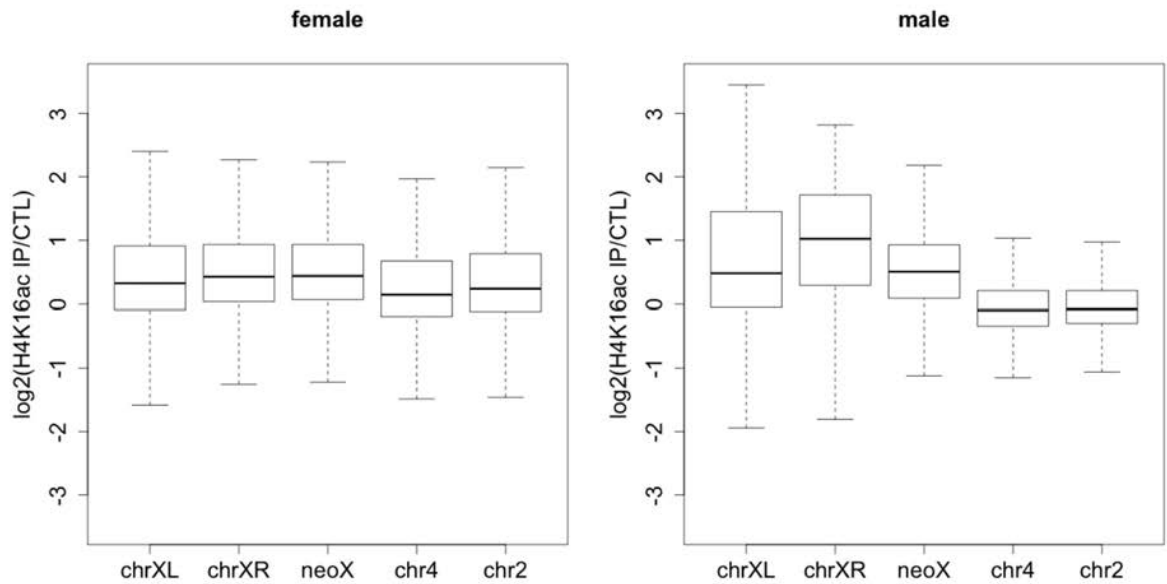

**Supplementary Figure 10. Chromatin structure of sex chromosomes vs. autosomes in males vs. females**. Chromatin structure (as measured by H4K16ac enrichment) is similar between the X and autosomes in females and differs dramatically on the X and autosomes in males of *D. miranda*. Each boxplot shows log2 read depth ratio of ChIP-seq vs. input control along the gene body including the flanking 3kb regions on a specific chromosome.

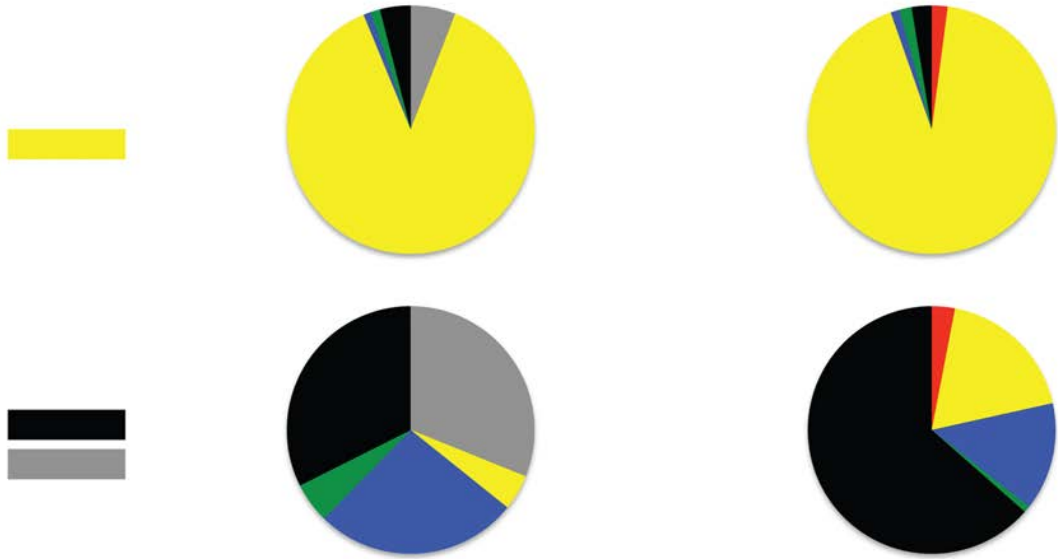

**Supplementary Figure 11. Chromatin states are overall conserved between *D. melanogaster* and *D. miranda* females.** We define chromatin types either in *D. melanogaster*, using the classification of [43], or in *D. miranda*, using the classification described in Supplementary Figure 8. The pie charts show the composition of a particular type of chromatin defined in one species (active 'yellow' chromatin on top; inactive 'black' [and 'grey' for *D. miranda*] on the bottom) in the other species. For example, the upper left pie shows the 'yellow' genes defined by *D. melanogaster* and their chromatin type compositions defined using *D. miranda* female data. Overall, both definitions of active vs. repressive chromatin show a high overlap between species, suggesting chromatin types of orthologous genes are relatively conserved.

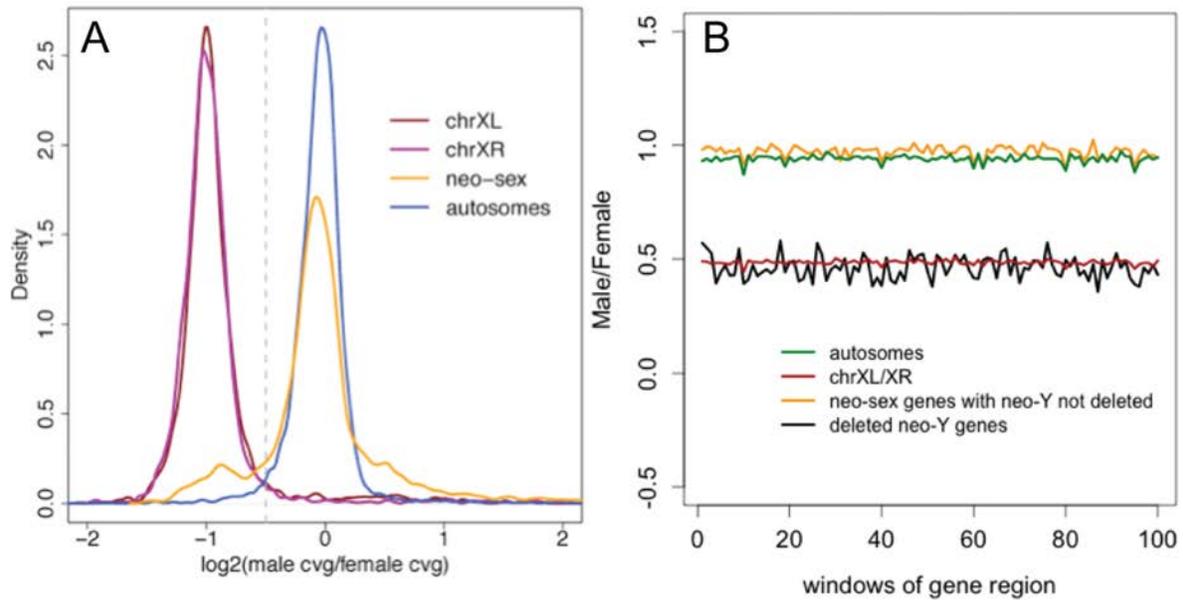

**Supplementary Figure 12. Identification of deleted genes on the neo-Y chromosome. A.** Shown is the histogram of male vs. female coverage ratios at exonic regions for all *D. miranda* genes. A cutoff (dotted line, log2(male/female)=-0.5) separating the distribution of autosomes and X chromosomes was picked to identify genes that are deleted from the neo-Y chromosome. **B.** Metagene plot of male/female coverage for different classes of genes (X-linked, autosomal, neo-sex genes with/without deleted neo-Y), across the gene body.

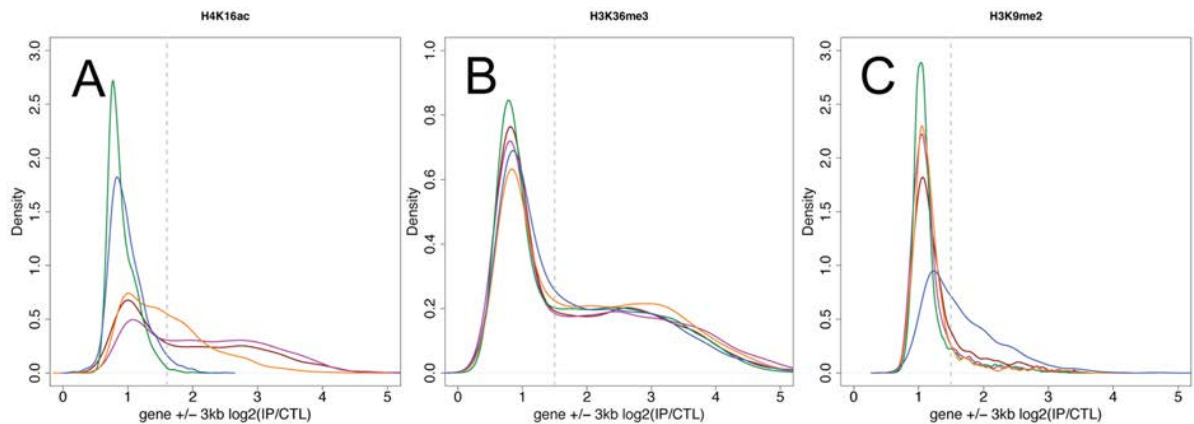

**Supplementary Figure 13. Definition of bound/unbound genes for different chromatin marks.** Shown is the histogram of the log2 coverage ratio of ChIP-seq vs. input control along the gene body including up/downstream 3kb regions separately for each chromosome. Autosomes are in green, chrXL in red, chrXR in purple, neo-X in orange and neo-Y in blue. Cutoffs discriminating bound/unbound genes were chosen where the bimodal distribution is separated for two peaks or sex/neo-sex chromosomes are separated from the autosomes.

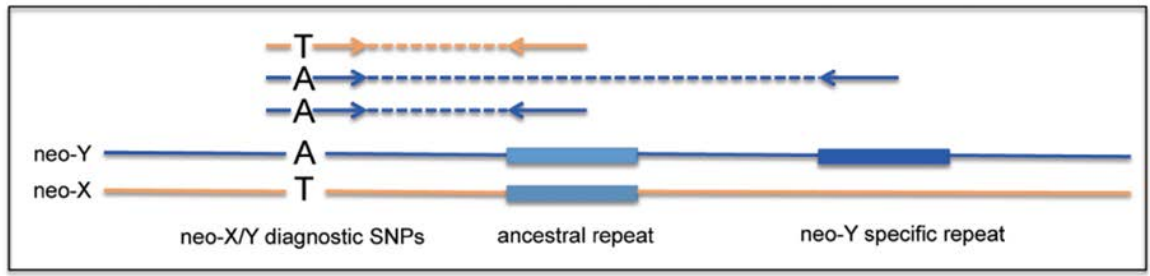

**Supplementary Figure 14. Schematic diagram of repeat enrichment analyses**. To identify neo-Y specific enrichment of repeat sequences, we counted the ratio of mate-pairs where one read spanned a neo-X/Y diagnostic SNPs and the other read mapped to a repeat sequence in our consensus repeat library for *D. miranda*

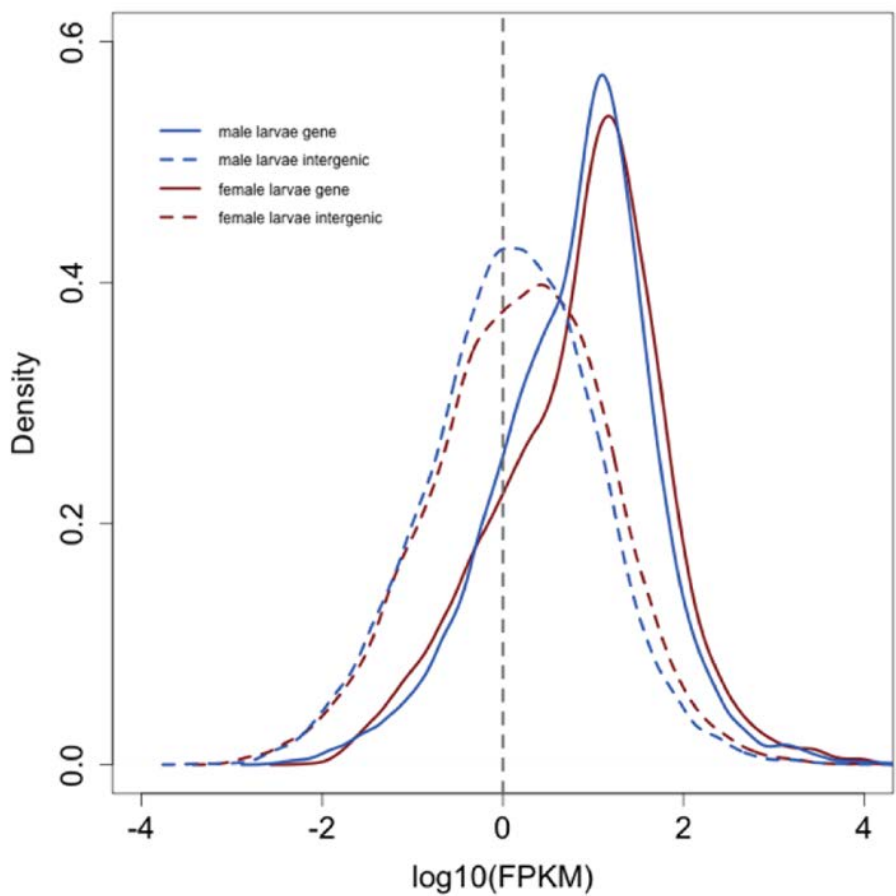

**Supplementary Figure 15. Identification of active and silent neo-Y genes.** Shown is the histogram of FPKM values derived from genes (solid line) and intergenic regions (dotted line). The peak of the FPKM distribution at intergenic regions is chosen as a cut-off to determine whether a gene is active or silent on the neo-Y.

**Supplementary Table 1.** MSL-binding and H4K16ac enrichment for genes on different chromosomes.

| | MSL+/ H4K16ac- | MSL+/ H4K16+ | MSL-/ H4K16ac+ | dosage compensated | MSL-/H4K16ac- (not compensated) | total |
|---|---|---|---|---|---|---|
| chrXL | 28 | 766 | 389 | 1183 | 1090 | 2273 |
| chrXR | 34 | 1238 | 774 | 2046 | 1113 | 3159 |
| neo-X | 62 | 545 | 596 | 1203 | 1506 | 2709 |
| autosomes | 0 | 0 | 52 | 52 | 6627 | 6679 |

**Supplementary Table 2**. GO terms significantly enriched in neo-X genes

| GO term id | GO domain | GO term name |
|---|---|---|
| GO terms significantly enriched in dosage compensated neo-X genes | | |
| GO:0032502 | biological_process | developmental_process |
| GO:0044707 | biological_process | single-multicellular_organism_process |
| GO:0051603 | biological_process | proteolysis_involved_in_cellular_protein_catabolic_process |
| GO:0019953 | biological_process | sexual_reproduction |
| GO:0006032 | biological_process | chitin_catabolic_process |
| GO:0007017 | biological_process | microtubule-based_process |
| GO:1901575 | biological_process | organic_substance_catabolic_process |
| GO:0030163 | biological_process | protein_catabolic_process |
| GO:0046716 | biological_process | muscle_cell_homeostasis |
| GO:0009057 | biological_process | macromolecule_catabolic_process |
| GO:0009056 | biological_process | catabolic_process |
| GO:0009987 | biological_process | cellular_process |
| GO:0044446 | cellular_component | intracellular_organelle_part |
| GO:0032991 | cellular_component | macromolecular_complex |
| GO:0005623 | cellular_component | cell |
| GO:0044422 | cellular_component | organelle_part |
| GO:0043226 | cellular_component | organelle |
| GO:0044464 | cellular_component | cell_part |
| GO:0001882 | molecular_function | nucleoside_binding |
| GO:0008135 | molecular_function | translation_factor_activity,_nucleic_acid_binding |
| GO terms significantly enriched in genes not dosage compensated on the neo-X | | |
| GO:0050794 | biological_process | regulation_of_cellular_process |
| GO:0060255 | biological_process | regulation_of_macromolecule_metabolic_process |
| GO:0044699 | biological_process | single-organism_process |
| GO:0050896 | biological_process | response_to_stimulus |
| GO:0019438 | biological_process | aromatic_compound_biosynthetic_process |
| GO:0009889 | biological_process | regulation_of_biosynthetic_process |
| GO:0051171 | biological_process | regulation_of_nitrogen_compound_metabolic_process |
| GO:0006508 | biological_process | proteolysis |
| GO:0010468 | biological_process | regulation_of_gene_expression |
| GO:0018130 | biological_process | heterocycle_biosynthetic_process |
| GO:0031326 | biological_process | regulation_of_cellular_biosynthetic_process |
| GO:0080090 | biological_process | regulation_of_primary_metabolic_process |
| GO:0015837 | biological_process | amine_transport |
| GO:0034654 | biological_process | nucleobase-containing_compound_biosynthetic_process |
| GO:0032774 | biological_process | RNA_biosynthetic_process |
| GO:0010556 | biological_process | regulation_of_macromolecule_biosynthetic_process |
| GO:0031323 | biological_process | regulation_of_cellular_metabolic_process |
| GO:0032501 | biological_process | multicellular_organismal_process |
| GO:0016070 | biological_process | RNA_metabolic_process |
| GO:0055085 | biological_process | transmembrane_transport |
| GO:0030182 | biological_process | neuron_differentiation |
| GO:0050789 | biological_process | regulation_of_biological_process |
| GO:0044271 | biological_process | cellular_nitrogen_compound_biosynthetic_process |
| GO:0006351 | biological_process | transcription,_DNA-dependent |
| GO:0030030 | biological_process | cell_projection_organization |
| GO:0019222 | biological_process | regulation_of_metabolic_process |
| GO:1901362 | biological_process | organic_cyclic_compound_biosynthetic_process |
| GO:2000112 | biological_process | regulation_of_cellular_macromolecule_biosynthetic_process |
| GO:0065007 | biological_process | biological_regulation |
| GO:0016020 | cellular_component | membrane |
| GO:0044425 | cellular_component | membrane_part |
| GO:0043234 | cellular_component | protein_complex |
| GO:0022892 | molecular_function | substrate-specific_transporter_activity |
| GO:0005549 | molecular_function | odorant_binding |
| GO:0005372 | molecular_function | water_transmembrane_transporter_activity |

**Supplementary Table 3.** ChIP-seq reads mapped to neo-X and neo-Y specific variants, and undifferentiated neo-sex linked regions.

|              | H4K16ac- | H3K9me3 | Input   |
|--------------|----------|---------|---------|
| neo-X        | 2616959  | 1898318 | 1623828 |
| neo-Y        | 781891   | 1221193 | 762854  |
| common       | 1327023  | 1782427 | 998135  |
| total mapped | 4725873  | 4901938 | 3384817 |

**Supplementary Data S1.** Expression values (FPKM) for genes on element C in *D. miranda* and *D. pseudoobscura* male and female larvae, and enrichment levels (log2[ChIP-seq / input control]) for MSL3, H4K16ac and H3K36me3 on the neo-X and H3K9me2 on the neo-Y of *D. miranda* males.